%Paper: hep-th/9207081
%From: Kanehisa Takasaki <TAKASAKI%JPNYITP.BITNET@pucc.princeton.edu>
%Date: Fri, 24 Jul 92 17:28:55 JST
%Date (revised): Wed, 03 Feb 93 11:27:23 JST

%%%%% Quasi-classical limit of KP hierarchy,
%%%%% W-symmetries and free fermions
%%%%% By Kanehisa Takasaki and Takashi Takebe, KUCP-0050/92
%%%%%%%%%%% macros extracted from "vanilla.sty" %%%%%%%%%%%%%%%%%%%%%%
\catcode`\@=11
\font\tensmc=cmcsc10      %change to CM fonts 3-31-87
%\font\tensmc=amcsc10
\def\smc{\tensmc}

\def\wlog#1{}
\newif\iftitle@
\outer\def\title{\title@true\vglue 24\p@ plus 12\p@ minus 12\p@
   \bgroup\let\\=\cr\tabskip\centering
   \halign to \hsize\bgroup\tenbf\hfill\ignorespaces##\unskip\hfill\cr}
\def\endtitle{\cr\egroup\egroup\vglue 18\p@ plus 12\p@ minus 6\p@}
\outer\def\author{\iftitle@\vglue -18\p@ plus -12\p@ minus -6\p@\fi\vglue
    12\p@ plus 6\p@ minus 3\p@\bgroup\let\\=\cr\tabskip\centering
    \halign to \hsize\bgroup\smc\hfill\ignorespaces##\unskip\hfill\cr}
\def\endauthor{\cr\egroup\egroup\vglue 18\p@ plus 12\p@ minus 6\p@}
\outer\def\heading{\bigbreak\bgroup\let\\=\cr\tabskip\centering
    \halign to \hsize\bgroup\smc\hfill\ignorespaces##\unskip\hfill\cr}
\def\endheading{\cr\egroup\egroup\nobreak\medskip}

\outer\def\endproclaim{\par\ifdim\lastskip<\medskipamount\removelastskip
  \penalty 55 \fi\medskip\rm}
\outer\def\demo#1{\par\ifdim\lastskip<\smallskipamount\removelastskip
    \smallskip\fi\noindent{\smc\ignorespaces#1\unskip:\enspace}\rm
      \ignorespaces}

\newcount\footmarkcount@
\footmarkcount@=1
\def\makefootnote@#1#2{\insert\footins{\interlinepenalty=100
  \splittopskip=\ht\strutbox \splitmaxdepth=\dp\strutbox
  \floatingpenalty=\@MM
  \leftskip=\z@\rightskip=\z@\spaceskip=\z@\xspaceskip=\z@
  \noindent{#1}\footstrut\rm\ignorespaces #2\strut}}
\def\footnote{\let\@sf=\empty\ifhmode\edef\@sf{\spacefactor
   =\the\spacefactor}\/\fi\futurelet\next\footnote@}
\def\footnote@{\ifx"\next\let\next\footnote@@\else
    \let\next\footnote@@@\fi\next}
\def\footnote@@"#1"#2{#1\@sf\relax\makefootnote@{#1}{#2}}
\def\footnote@@@#1{$^{\number\footmarkcount@}$\makefootnote@
   {$^{\number\footmarkcount@}$}{#1}\global\advance\footmarkcount@ by 1 }
\def\eat@#1{}
\mathchardef\prime@="0230
\def\prime{{{}\prime@{}}}
\def\prim@s{\prime@\futurelet\next\pr@m@s}

\def\,{\relax\ifmmode\mskip\thinmuskip\else\thinspace\fi}
\def\!{\relax\ifmmode\mskip-\thinmuskip\else\negthinspace\fi}
\def\frac#1#2{{#1\over#2}}
\def\dfrac#1#2{{\displaystyle{#1\over#2}}}

\def\:{\nobreak\hskip.1111em{:}\hskip.3333em plus .0555em\relax}
\def\intic@{\mathchoice{\hskip5\p@}{\hskip4\p@}{\hskip4\p@}{\hskip4\p@}}
\def\negintic@
 {\mathchoice{\hskip-5\p@}{\hskip-4\p@}{\hskip-4\p@}{\hskip-4\p@}}
\def\intkern@{\mathchoice{\!\!\!}{\!\!}{\!\!}{\!\!}}
\def\intdots@{\mathchoice{\cdots}{{\cdotp}\mkern1.5mu
    {\cdotp}\mkern1.5mu{\cdotp}}{{\cdotp}\mkern1mu{\cdotp}\mkern1mu
      {\cdotp}}{{\cdotp}\mkern1mu{\cdotp}\mkern1mu{\cdotp}}}
\newcount\intno@
\def\iint{\intno@=\tw@\futurelet\next\ints@}
\def\iiint{\intno@=\thr@@\futurelet\next\ints@}
\def\iiiint{\intno@=4 \futurelet\next\ints@}
\def\idotsint{\intno@=\z@\futurelet\next\ints@}
\def\ints@{\findlimits@\ints@@}
\newif\iflimtoken@
\newif\iflimits@
\def\findlimits@{\limtoken@false\limits@false\ifx\next\limits
 \limtoken@true\limits@true
   \else\ifx\next\nolimits\limtoken@true\limits@false
    \fi\fi}
\def\multintlimits@{\intop\ifnum\intno@=\z@\intdots@
  \else\intkern@\fi
    \ifnum\intno@>\tw@\intop\intkern@\fi
     \ifnum\intno@>\thr@@\intop\intkern@\fi\intop}
\def\multint@{\int\ifnum\intno@=\z@\intdots@\else\intkern@\fi
   \ifnum\intno@>\tw@\int\intkern@\fi
    \ifnum\intno@>\thr@@\int\intkern@\fi\int}
\def\ints@@{\iflimtoken@\def\ints@@@{\iflimits@
   \negintic@\mathop{\intic@\multintlimits@}\limits\else
    \multint@\nolimits\fi\eat@}\else
     \def\ints@@@{\multint@\nolimits}\fi\ints@@@}
\def\Sb{_\bgroup\vspace@
        \baselineskip=\fontdimen10 \scriptfont\tw@
        \advance\baselineskip by \fontdimen12 \scriptfont\tw@
        \lineskip=\thr@@\fontdimen8 \scriptfont\thr@@
        \lineskiplimit=\thr@@\fontdimen8 \scriptfont\thr@@
        \Let@\vbox\bgroup\halign\bgroup \hfil$\scriptstyle
            {##}$\hfil\cr}
\def\endSb{\crcr\egroup\egroup\egroup}
\def\Sp{^\bgroup\vspace@
        \baselineskip=\fontdimen10 \scriptfont\tw@
        \advance\baselineskip by \fontdimen12 \scriptfont\tw@
        \lineskip=\thr@@\fontdimen8 \scriptfont\thr@@
        \lineskiplimit=\thr@@\fontdimen8 \scriptfont\thr@@
        \Let@\vbox\bgroup\halign\bgroup \hfil$\scriptstyle
            {##}$\hfil\cr}
\def\endSp{\crcr\egroup\egroup\egroup}
\def\Let@{\relax\iffalse{\fi\let\\=\cr\iffalse}\fi}
\def\vspace@{\def\vspace##1{\noalign{\vskip##1 }}}
\def\aligned{\,\vcenter\bgroup\vspace@\Let@\openup\jot\m@th\ialign
  \bgroup \strut\hfil$\displaystyle{##}$&$\displaystyle{{}##}$\hfil\crcr}
\def\endaligned{\crcr\egroup\egroup}
\def\matrix{\,\vcenter\bgroup\Let@\vspace@
    \normalbaselines
  \m@th\ialign\bgroup\hfil$##$\hfil&&\quad\hfil$##$\hfil\crcr
    \mathstrut\crcr\noalign{\kern-\baselineskip}}
\def\endmatrix{\crcr\mathstrut\crcr\noalign{\kern-\baselineskip}\egroup
                \egroup\,}
\newtoks\hashtoks@
\hashtoks@={#}
\def\format{\crcr\egroup\iffalse{\fi\ifnum`}=0 \fi\format@}
\def\format@#1\\{\def\preamble@{#1}%
  \def\c{\hfil$\the\hashtoks@$\hfil}%
  \def\r{\hfil$\the\hashtoks@$}%
  \def\l{$\the\hashtoks@$\hfil}%
  \setbox\z@=\hbox{\xdef\Preamble@{\preamble@}}\ifnum`{=0 \fi\iffalse}\fi
   \ialign\bgroup\span\Preamble@\crcr}

\def\cases{\left\{\,\vcenter\bgroup\vspace@
     \normalbaselines\openup\jot\m@th
       \Let@\ialign\bgroup$##$\hfil&\quad$##$\hfil\crcr
      \mathstrut\crcr\noalign{\kern-\baselineskip}}

\newif\iftagsleft@
\tagsleft@true
\def\TagsOnRight{\global\tagsleft@false}
\def\tag#1$${\iftagsleft@\leqno\else\eqno\fi
 \hbox{\def\pagebreak{\global\postdisplaypenalty-\@M}%
 \def\nopagebreak{\global\postdisplaypenalty\@M}\rm(#1\unskip)}%
  $$\postdisplaypenalty\z@\ignorespaces}
\interdisplaylinepenalty=\@M
\def\allowdisplaybreak@{\def\allowdisplaybreak{\noalign{\allowbreak}}}
\def\displaybreak@{\def\displaybreak{\noalign{\break}}}
\def\align#1\endalign{\def\tag{&}\vspace@\allowdisplaybreak@\displaybreak@
  \iftagsleft@\lalign@#1\endalign\else
   \ralign@#1\endalign\fi}
\def\ralign@#1\endalign{\displ@y\Let@\tabskip\centering
   \halign to\displaywidth
     {\hfil$\displaystyle{##}$\tabskip=\z@&$\displaystyle{{}##}$\hfil
       \tabskip=\centering&\llap{\hbox{(\rm##\unskip)}}\tabskip\z@\crcr
             #1\crcr}}
\def\lalign@
 #1\endalign{\displ@y\Let@\tabskip\centering\halign to \displaywidth
   {\hfil$\displaystyle{##}$\tabskip=\z@&$\displaystyle{{}##}$\hfil
   \tabskip=\centering&\kern-\displaywidth
        \rlap{\hbox{(\rm##\unskip)}}\tabskip=\displaywidth\crcr
               #1\crcr}}
\def\overrightarrow{\mathpalette\overrightarrow@}
\def\overrightarrow@#1#2{\vbox{\ialign{$##$\cr
    #1{-}\mkern-6mu\cleaders\hbox{$#1\mkern-2mu{-}\mkern-2mu$}\hfill
     \mkern-6mu{\to}\cr
     \noalign{\kern -1\p@\nointerlineskip}
     \hfil#1#2\hfil\cr}}}
\def\overleftarrow{\mathpalette\overleftarrow@}
\def\overleftarrow@#1#2{\vbox{\ialign{$##$\cr
     #1{\leftarrow}\mkern-6mu\cleaders
      \hbox{$#1\mkern-2mu{-}\mkern-2mu$}\hfill
      \mkern-6mu{-}\cr
     \noalign{\kern -1\p@\nointerlineskip}
     \hfil#1#2\hfil\cr}}}
\def\overleftrightarrow{\mathpalette\overleftrightarrow@}
\def\overleftrightarrow@#1#2{\vbox{\ialign{$##$\cr
     #1{\leftarrow}\mkern-6mu\cleaders
       \hbox{$#1\mkern-2mu{-}\mkern-2mu$}\hfill
       \mkern-6mu{\to}\cr
    \noalign{\kern -1\p@\nointerlineskip}
      \hfil#1#2\hfil\cr}}}
\def\underrightarrow{\mathpalette\underrightarrow@}
\def\underrightarrow@#1#2{\vtop{\ialign{$##$\cr
    \hfil#1#2\hfil\cr
     \noalign{\kern -1\p@\nointerlineskip}
    #1{-}\mkern-6mu\cleaders\hbox{$#1\mkern-2mu{-}\mkern-2mu$}\hfill
     \mkern-6mu{\to}\cr}}}
\def\underleftarrow{\mathpalette\underleftarrow@}
\def\underleftarrow@#1#2{\vtop{\ialign{$##$\cr
     \hfil#1#2\hfil\cr
     \noalign{\kern -1\p@\nointerlineskip}
     #1{\leftarrow}\mkern-6mu\cleaders
      \hbox{$#1\mkern-2mu{-}\mkern-2mu$}\hfill
      \mkern-6mu{-}\cr}}}
\def\underleftrightarrow{\mathpalette\underleftrightarrow@}
\def\underleftrightarrow@#1#2{\vtop{\ialign{$##$\cr
      \hfil#1#2\hfil\cr
    \noalign{\kern -1\p@\nointerlineskip}
     #1{\leftarrow}\mkern-6mu\cleaders
       \hbox{$#1\mkern-2mu{-}\mkern-2mu$}\hfill
       \mkern-6mu{\to}\cr}}}
\def\sqrt#1{\radical"270370 {#1}}
\def\dots{\relax\ifmmode\let\next=\ldots\else\let\next=\tdots@\fi\next}
\def\tdots@{\unskip\ \tdots@@}
\def\tdots@@{\futurelet\next\tdots@@@}
\def\tdots@@@{$\mathinner{\ldotp\ldotp\ldotp}\,
   \ifx\next,$\else
   \ifx\next.\,$\else
   \ifx\next;\,$\else
   \ifx\next:\,$\else
   \ifx\next?\,$\else
   \ifx\next!\,$\else
   $ \fi\fi\fi\fi\fi\fi}
\def\text{\relax\ifmmode\let\next=\text@\else\let\next=\text@@\fi\next}
\def\text@@#1{\hbox{#1}}
\def\text@#1{\mathchoice
 {\hbox{\everymath{\displaystyle}\def\textfonti{\the\textfont1 }%
    \def\textfontii{\the\textfont2 }\textdef@@ T#1}}
 {\hbox{\everymath{\textstyle}\def\textfonti{\the\textfont1 }%
    \def\textfontii{\the\textfont2 }\textdef@@ T#1}}
 {\hbox{\everymath{\scriptstyle}\def\textfonti{\the\scriptfont1 }%
   \def\textfontii{\the\scriptfont2 }\textdef@@ S\rm#1}}
 {\hbox{\everymath{\scriptscriptstyle}%
   \def\textfonti{\the\scriptscriptfont1 }%
   \def\textfontii{\the\scriptscriptfont2 }\textdef@@ s\rm#1}}}
\def\textdef@@#1{\textdef@#1\rm \textdef@#1\bf
   \textdef@#1\sl \textdef@#1\it}

\def\textdef@#1#2{%
 \def\next{\csname\expandafter\eat@\string#2fam\endcsname}%
\if S#1\edef#2{\the\scriptfont\next\relax}%
 \else\if s#1\edef#2{\the\scriptscriptfont\next\relax}%
 \else\edef#2{\the\textfont\next\relax}\fi\fi}
\scriptfont\itfam=\tenit \scriptscriptfont\itfam=\tenit
\scriptfont\slfam=\tensl \scriptscriptfont\slfam=\tensl
\mathcode`\0="0030
\mathcode`\1="0031
\mathcode`\2="0032
\mathcode`\3="0033
\mathcode`\4="0034
\mathcode`\5="0035
\mathcode`\6="0036
\mathcode`\7="0037
\mathcode`\8="0038
\mathcode`\9="0039
\def\Cal{\relax\ifmmode\let\next=\Cal@\else
    \def\next{\errmessage{Use \string\Cal\space only in %
      math mode}}\fi\next}
    \def\Cal@#1{{\fam2 #1}}
\def\bold{\relax\ifmmode\let\next=\bold@\else
    \def\next{\errmessage{Use \string\bold\space only in %
      math mode}}\fi\next}
    \def\bold@#1{{\fam\bffam #1}}
\mathchardef\Gamma="0000
\mathchardef\Delta="0001
\mathchardef\Theta="0002
\mathchardef\Lambda="0003
\mathchardef\Xi="0004
\mathchardef\Pi="0005
\mathchardef\Sigma="0006
\mathchardef\Upsilon="0007
\mathchardef\Phi="0008
\mathchardef\Psi="0009
\mathchardef\Omega="000A
\mathchardef\varGamma="0100
\mathchardef\varDelta="0101
\mathchardef\varTheta="0102
\mathchardef\varLambda="0103
\mathchardef\varXi="0104
\mathchardef\varPi="0105
\mathchardef\varSigma="0106
\mathchardef\varUpsilon="0107
\mathchardef\varPhi="0108
\mathchardef\varPsi="0109
\mathchardef\varOmega="010A
\def\wlog#1{\immediate\write-1{#1}}
%%\catcode`\@=\active
\catcode`\@=12  %% defining '@' as a letter
%%%%%%%%%%%%%%%%%%%%%%%%%%%%%%%%%%%%%%%%%%%%%%%%%%%%%%%%%%%%%%%%%%
\mag=\magstep1
\baselineskip=20pt
\hsize=16truecm
\vsize=23truecm
\TagsOnRight
\overfullrule=0pt
\def\=def{\; \mathop{=}_{\text{\rm def}} \;}
\def\rd{\partial}
\def\Res{\; \mathop{\text{\rm Res}} \;}

\def\bfZ{{\bold Z}}
\def\calB{{\cal B}}
\def\calL{{\cal L}}
\def\calM{{\cal M}}

\def\calO{{\cal O}}

\def\rdtilde{\tilde{\rd}}
\def\lambdatilde{\tilde{\lambda}}

\def\SDiff(2){ \text{SDiff(2)} }

%%%%% cover %%%%%%%%%%%%%%%%%%%%%%%%%%%%%%%%%%%%%%%%%%%%%%%%%%%%%%%%%%
\pageno=1
\line{{\it College of Liberal Arts and Sciences}
  \hfill KUCP-0050/92}
\line{{\it Kyoto University}
  \hfill July 1992}

\title
    \smc Quasi-classical limit of KP hierarchy, \\
    \smc W-symmetries and free fermions \\
\endtitle

\author
    Kanehisa Takasaki$^1$ and Takashi Takebe$^2$\\
    \\
    $^1$%%
    {\rm Institute of Mathematics, Yoshida College, Kyoto University}\\
    {\rm Yoshida-Nihonmatsu-cho, Sakyo-ku, Kyoto 606, Japan}\\
    {\rm E-mail: takasaki @ jpnyitp (Bitnet)}\\
    \\
    $^2$%%
    {\rm Department of Mathematical Sciences, University of Tokyo}\\
    {\rm Hongo, Bunkyo-ku, Tokyo 113, Japan}\\
    {\rm E-mail: takebe @ math.s.u-tokyo.ac.jp}\\
\endauthor

\title Abstract
\endtitle
\noindent
This paper deals with the dispersionless KP hierarchy from the point
of view of quasi-classical limit. Its Lax formalism, W-infinity
symmetries and general solutions are shown to be reproduced from
their counterparts in the KP hierarchy in the limit of $\hbar \to 0$.
Free fermions and bosonized vertex operators play a key role in
the description of W-infinity symmetries and general solutions,
which is technically very similar to a recent free fermion formalism
of $c=1$ matrix models.

\vfill
\eject
%%%%% text %%%%%%%%%%%%%%%%%%%%%%%%%%%%%%%%%%%%%%%%%%%%%%%%%%%%%%%

Dispersionless versions of integrable hierarchies (KP, Toda, etc.)
[1], like the ordinary hierarchies, have many applications in low
dimensional integrable field theories. For example, the dispersionless
KP hierarchy, as well as more general hierarchies of Whitham-type
[2,3], possesses many solutions that are related to topological
conformal field theory [4,5]. The dispersionless Toda hierarchy
arises in continuous (or large-$N$) limit of the ordinary Toda field
theory [6-8], and also related to 4D self-dual gravity [9] via
dimensional reduction.

In previous work on the dispersionless KP and Toda hierarchies [10],
we stressed analogy of these hierarchies with self-dual gravity.
A main result, which we believe reflects the integrable nature
of these hierarchies most clearly, is the existence of infinitely
many symmetries with the structure of a W-infinity algebra,
$w_{1+\infty}$ [11]. Most ideas and tools used therein are borrowed
from a construction of similar $\calL w_{1+\infty}$ (the loop algebra
of $w_{1+\infty}$) symmetries of self-dual gravity in earlier papers
[12,13].

This approach, in its nature, tells us nothing on how the
$w_{1+\infty}$ symmetries of a dispersionless hierarchy are
related to $W_{1+\infty}$ symmetries of the original hierarchy,
i.e., symmetries with the structure of another typical W-infinity
algebra, $W_{1+\infty}$ [11].  These $W_{1+\infty}$ symmetries
are known to play a key role in 2D quantum and topological gravity
[14-18] as ``W-constraints." Actually, their $w_{1+\infty}$ versions
also emerge in similar models [2,10]. The two types of W-infinity
symmetries should be related more directly!

In this paper, we reorganize our previous work along with new results
from the point of view of quasi-classical limit. Dispersionless
limit, in fact, can also be understood as quasi-classical limit, as
already well recognized in 2D gravity [17]. For illustration,
we mostly deal with the dispersionless KP hierarchy, and derive
its Lax formalism, tau function, W-infinity symmetries, and general
solutions from their counterparts in the KP hierarchy in the limit
of $\hbar \to 0$.

The dispersionless KP hierarchy (which we called SDiff(2) KP hierarchy
in the previous paper [10]) has a Lax-type representation [1-3]
$$
    \frac{\rd \calL}{\rd t_n} = \{ \calB_n, \calL \}, \quad
    \calB_n \=def (\calL^n)_{\ge 0}, \quad
    n = 1,2,\ldots,                                   \tag 1
$$
where $\calL$ is a Laurent series in the ``momentum'' $k$ (conjugate
variable of $x = t_1$) of the form
$$
    \calL = k + \sum_{n=1}^\infty u_{n+1}(t) k^{-n},  \tag 2
$$
``$(\quad)_{\ge 0}$'' means the projection onto a polynomial in
$k$ dropping negative powers, and ``$\{\quad,\quad\}$''
the Poisson bracket in 2D ``phase space'' $(k,x)$,
$$
    \{ A(k,x), B(k,x) \}
    =  \frac{\rd A(k,x)}{\rd k}\frac{\rd B(k,x)}{\rd x}
      -\frac{\rd A(k,x)}{\rd x}\frac{\rd B(k,x)}{\rd k}.
                                                      \tag 3
$$
This obviously imitates the Lax representation of the KP hierarchy,
replacing pseudo-differential operators (in $x$) and their commutators
by Laurent series (in $k$) and Poisson brackets. We argued in the
previous work [10] that this Lax representation should be extended
to a larger system. The extended Lax representation consists of
$\calL$ and another Laurent series $\calM$ of the form
$$
    \calM = \sum_{n=1}^\infty nt_n \calL^{n-1}
          + \sum_{n=1}^\infty v_{n+1}(t) \calL^{-n-1}  \tag 4
$$
that satisfies the same Lax-type equations and the canonical Poisson
relation:
$$
\align
  & \frac{\rd \calM}{\rd t_n} = \{ \calB_n, \calM \}, \quad
    n = 1,2,\ldots,                                    \tag 5 \\
  & \{ \calL, \calM \} = 1.                            \tag 6 \\
\endalign
$$
Actually, the above three sets of equations for $\calL$ and $\calM$ can
be cast into a single exterior differential equation,
$$
    d\calL \wedge d\calM = \sum_{n=1}^\infty d\calB_n \wedge dt_n,
                                                       \tag 7
$$
which is very similar to a 2-form equation that lies in the heart of
integrability of self-dual gravity [12,13]; our previous construction
of $w_{1+\infty}$ is based on this fact. An analogue of the KP tau
function, which we here write $\tau_{\text{dKP}}(t)$, is also
introduced in our previous work as:
$$
    \frac{ \rd \log\tau_{\text{dKP}} }{\rd t_n} = v_{n+1}, \quad
    n=1,2,\ldots.                                      \tag 8
$$

We now introduce a ``Planck constant" $\hbar$ into the ordinary
setting of the KP hierarchy [19-21]. As in the dispersionless
case, it is crucial to extend the ordinary Lax formalism by adding
a second Lax operator $M$ [22,16-18]. The extended Lax
representation of the KP hierarchy, in the presence of $\hbar$,
is given by
$$
\align
  & \hbar \frac{\rd L}{\rd t_n} = [ B_n, L ], \quad
    B_n = (L^n)_{\ge 0},                                      \\
  & \hbar \frac{\rd M}{\rd t_n} = [ B_n, M ], \quad
    n=1,2,\ldots,                                             \\
  & [ L, M ] = \hbar,                                  \tag 9 \\
\endalign
$$
where $L$ and $M$, counterparts of $\calL$ and $\calM$, are
pseudo-differential operators (in $x=t_1$),
$$
\align
  & L = \hbar \rd
      + \sum_{n=1}^\infty u_{n+1}(\hbar,t)(\hbar\rd)^{-n}, \quad
    \rd = \rd/\rd x,                                           \\
  & M = \sum_{n=1}^\infty n t_n L^{n-1}
      + \sum_{n=1}^\infty v_{n+1}(\hbar,t) L^{-n-1},   \tag 10 \\
\endalign
$$
and ``$(\quad)_{\ge 0}$'' now stands for the projection onto a
differential operator dropping negative powers of $\rd$. The
coefficients $u_n(\hbar,t)$ and $v_n(\hbar,t)$ are assumed to
have such an asymptotic form as $u_n(\hbar,t) = u_n(t) + O(\hbar)$,
$v_n(\hbar,t) = v_n(t) + O(\hbar)$ in the limit of $\hbar \to 0$.
Accordingly the Baker-Akhiezer function and the tau function, too,
are to depend on $\hbar$.  The Baker-Akhiezer function
$\Psi(\hbar,t,\lambda)$, where $\lambda$ is the so called ``spectral
parameter,'' is by definition a function (or a formal Laurent series
of $\lambda$) with the asymptotic form
$$
    \Psi(\hbar,t,\lambda)
    = ( 1 + O(\lambda^{-1}) )
      \exp ( \hbar^{-1} \sum_{n=1}^\infty t_n \lambda^n) \quad
    (\lambda \to \infty, \ \hbar \to 0)                \tag 11
$$
that satisfies the linear equations
$$
    \lambda \Psi = L \Psi, \quad
    \hbar \frac{\rd \Psi}{\rd \lambda} = M \Psi, \quad
    \hbar \frac{\rd \Psi}{\rd t_n} = B_n \Psi.         \tag 12
$$
The tau function $\tau(\hbar,t)$ is by definition a function
that reproduces the Baker-Akhiezer function as
$$
\align
  & \Psi(\hbar,t)
    = \dfrac{ \tau(\hbar, t - \hbar\epsilon(\lambda^{-1})) }
            { \tau(\hbar,t ) }
      \exp( \hbar^{-1} \sum_{n=1}^\infty t_n \lambda^n ),
                                                              \\
  &  \epsilon(\lambda^{-1})
    = \left( \frac{1}{\lambda}, \frac{1}{2\lambda^2},
             \frac{1}{3\lambda^3}, \ldots            \right).
                                                       \tag 13\\
\endalign
$$

The dispersionless KP hierarchy now emerges in the limit of
$\hbar \to 0$ as follows. Following the prescription of Kodama
and Gibbons [1], we rewrite the Baker-Akhiezer function into
the WKB form:
$$
    \Psi(\hbar,t,\lambda)
    = \exp[ \hbar^{-1} S(t,\lambda) + O(\hbar^0)] \quad
    (\hbar \to 0).                                       \tag 14
$$
The ``phase function'' $S(t,\lambda)$ then satisfies a set of
Hamilton-Jacobi (or eikonal) equations that follow from the
linear equations of the Baker-Akhiezer function. Those arising
from the second and third linear equations can be cast into a
single 1-form equation:
$$
    d S(t,\lambda) = \calM(t,\lambda) d \lambda
      + \sum_{n=1}^\infty \calB_n(t,\lambda) dt_n,       \tag 15
$$
where
$$
\align
  & \calM(t,\lambda)
      = \sum_{n=1}^\infty nt_n \lambda^{n-1}
       +\sum_{n=1}^\infty v_{n+1}(t) \lambda^{-n-1},
                                                                 \\
  & \calB_n(t,\lambda)
      = \left( \frac{\rd S(t,\lambda)}{\rd x}\right)^n
       +\sum_{i=0}^{n-2} b_{n,i}(t)
          \left( \frac{\rd S(t,\lambda)}{\rd x} \right)^i,
                                                         \tag 16 \\
\endalign
$$
and the coefficients $b_{n,i}(t)$, like $u_n(t)$ and $v_n(t)$,
are the leading ($\hbar^0$) term in $\hbar$-expansion of the
coefficients $b_{n,i}(\hbar,t)$ of
$B_n = (\hbar\rd)^n + \sum_{i=0}^{n-2}b_{n,i}(\hbar,t)(\hbar\rd)^i$.
The Hamilton-Jacobi equation for $\lambda\Psi = L\Psi$ is given by
$$
    \lambda = \frac{\rd S(t,\lambda)}{\rd x}
             +\sum_{n=1}^\infty u_{n+1}(t)
               \left( \frac{\rd S(t,\lambda)}{\rd x} \right)^{-n}.
                                                         \tag 17
$$
If we make change of the independent variables
$(t,\lambda) \to (t,k)$ by
$$
    k = \frac{\rd S(t,\lambda)}{\rd x}
    \quad (x=t_1),                                       \tag 18
$$
$\lambda$ becomes a function of $(t,k)$ as
$\lambda = \calL(t,k) = k + \sum_{n=1}^\infty u_{n+1}(t) k^{-n}$,
and we can define $\calM = \calM(t,\calL)$ and
$\calB_n = \calB_n(t,\calL)$ in a desired form.
These functions $\calL$, $\calM$, and $\calB_n$ satisfy 2-form
equation (7) (hence the dispersionless KP hierarchy)
as an immediate consequence of 1-form equation (15).

One can easily show that the tau function $\tau(\hbar,t)$ has
to behave as
$$
    \tau(\hbar,t) = \exp[ \hbar^{-2}F(t) + O(\hbar^{-1})]  \quad
    (\hbar \to 0)                                        \tag 19
$$
so as to reproduce the WKB asymptotic form of the Baker-Ahkiezer
function. Here $F(t)$ is a function that satisfies the equations
$$
    \frac{\rd F(t)}{\rd t_n} = v_{n+1}(t), \quad
    n = 1,2,\ldots,
                                                         \tag 20
$$
therefore can be identified with $\log\tau_{\text{dKP}}$.
This relation resembles the relation between the partition
function and the free energy of matrix models in large-$N$ limit
($\hbar \sim 1/N$) [23]. Because of this, $F(t)$ may be called
the ``free energy'' of the dispersionless KP hierarchy.

We now turn to the issue of W-infinity symmetries. Basic notations
and tools are mostly borrowed from the work of Date et al. exploiting
free fermions and vertex operators [21]. We first modify their
bosonized vertex operator $Z(\lambdatilde,\lambda)$ by rescaling
$t_n \to \hbar^{-1}t_n$, $\rd/\rd t_n \to \hbar \rd/\rd t_n$ as:
$$
\align
  & Z(\hbar,\lambdatilde,\lambda)
    = \dfrac
       { \exp [\hbar^{-1} t(\lambdatilde) - \hbar^{-1} t(\lambda)]
         \exp[-\hbar \rdtilde(\lambdatilde^{-1})
              +\hbar \rdtilde(\lambda^{-1})]
         -1 }
       {\lambdatilde-\lambda},                                   \\
  & t(\lambda) = \sum_{n=1}^\infty t_n \lambda^n, \quad
    \rdtilde(\lambda^{-1}) = \sum_{n=1}^\infty
         \lambda^{-n} \frac{1}{n} \frac{\rd}{\rd t_n}.
                                                         \tag 21 \\
\endalign
$$
This rescaled vertex operator gives a 2-parameter family of
infinitesimal symmetries of the KP hierarchy with $\hbar$ inserted,
acting on the tau function as
$\tau \to \tau + \epsilon Z(\hbar,\lambdatilde,\lambda)\tau$.
A set of generators of $W_{1+\infty}$ symmetries,
$W^{(\ell)}_n(\hbar)$ ($\ell \ge 1$, $n \in \bfZ$), are
given by
$$
    W^{(\ell)}(\hbar,\lambda)
    = \sum_{n=-\infty}^\infty W^{(\ell)}_n(\hbar) \lambda^{-n-\ell}
    = \left( \frac{\rd}{\rd\lambdatilde} \right)^{\ell-1}
        Z(\hbar,\lambdatilde,\lambda)|_{\lambdatilde=\lambda}.
                                                             \tag 22
$$
Each of $W^{(\ell)}_n(\hbar)$ is a differential operator of finite
order. For example,
$$
\align
  & W^{(1)}_n(\hbar)    = \hbar \rd/\rd t_n,  \quad
    W^{(1)}_{-n}(\hbar) = nt_n/\hbar,         \quad
       n=1,2,\ldots,                                           \\
  & W^{(1)}_0(\hbar)    = 0,                         \tag 23   \\
\endalign
$$
and they form, along with the identity operator 1, a U(1)-current
(or Heisenberg) algebra.

These $W_{1+\infty}$ symmetries give rise to $w_{1+\infty}$ symmetries
of the dispersionless KP hierarchy as follows. Let us define
$w^{(\ell)}_n F$ by the limit
$$
    w^{(\ell)}_n F
    = \lim_{\hbar \to 0} \tau(\hbar,t)^{-1}
       \hbar^\ell W^{(\ell)}_n(\hbar) \tau(\hbar,t).
                                                      \tag 24
$$
Simple calculations show that this limit does exists and is given by
$$
    w^{(\ell)}_n F
    = \frac{1}{\ell} \Res \calM^\ell \calL^{n+\ell-1} d\calL,
                                                      \tag 25
$$
where ``$\Res$'' means the coefficient of $dk/k$ in a Laurent
series of $k$ (or, equivalently, the coefficients of
$d\calL/\calL$ in a Laurent series of $\calL$). These are nothing
but the generators of $w_{1+\infty}$ symmetries constructed in
our previous paper [10], which act on $F(t)$ as
$F \to F + \epsilon w^{(\ell)}_n F$.

To translate these observations into the fermionic language,
let us recall that the tau function of the ordinary ($\hbar=1$) KP
hierarchy can be written, in general, as a vacuum
expectation value,
$$
    \tau(t) = <0| e^{H(t)} g |0>.                     \tag 26
$$
Here the Fock vacuum states $<0|$ and $|0>$ are annihilated by
half of Fourier modes of the Date-Jimbo-Kashiwara-Miwa free
fermion fields
$$
    \psi(\lambda) = \sum_{n=-\infty}^\infty \psi_n \lambda^n, \quad
    \psi^*(\lambda) = \sum_{n=-\infty}^\infty \psi^*_n\lambda^{-n-1}
                                                      \tag 27
$$
with the anti-commutation relations
$$
    [ \psi_i, \psi_j ]_{+} = [\psi^*_i, \psi^*_j ]_{+} = 0, \quad
    [ \psi_i, \psi^*_j ]_{+} = \delta_{ij}.           \tag 28
$$
as
$$
\align
  & \psi_n |0>   = 0  \quad (n \le -1), \quad
    \psi^*_n |0> = 0  \quad (n \ge 0),                        \\
  & <0| \psi_n   = 0  \quad (n \ge 0), \quad
    <0| \psi^*_n = 0  \quad (n \le -1).              \tag 29  \\
\endalign
$$
The generator $H(t)$ of the time evolutions is given by
$$
    H(t) = \sum_{n=1}^\infty t_n H_n, \quad
    H_n = \sum_{i=-\infty}^\infty :\psi_i \psi^*_{i+n}:.
                                                     \tag 30
$$
Finally, $g$ is a Clifford operator written, in general, as
$$
    g = \text{const.} \exp \int\int : A(\lambdatilde,\lambda)
        \psi(\lambdatilde)\psi^*(\lambda): d\lambdatilde d\lambda.
                                                     \tag 31
$$
If the Planck constant $\hbar$ comes into the game, the above
expression of the tau function should change as:
$$
    \tau(\hbar,t) = <0| e^{H(t)/\hbar} g(\hbar) |0>. \tag 32
$$
The Clifford operator $g(\hbar)$ has to be suitably chosen
so as to ensure the correct asymptotic form of the tau function
in the limit of $\hbar \to 0$.

An ansatz for $g(\hbar)$ can be inferred as follows.  Let us
first note that $W_{1+\infty}$ symmetries are implemented by
inserting a fermion bilinear form in front of the Clifford
operator $g(\hbar)$.  This is due to the basic relation
$$
    Z(\hbar,\lambdatilde,\lambda)\tau(\hbar,t)
    = <0| e^{H(t)/\hbar} :\psi(\lambdatilde)\psi^*(\lambda):
          g(\hbar) |0>                               \tag 33
$$
pointed by Date et al. in the case of $\hbar=1$. For example,
insertion of
$$
    \calO^{(\ell)}_n(\hbar) =
    \oint :\lambda^{n+\ell-1}
           \left( \hbar \frac{\rd}{\rd\lambda}\right)^{\ell-1}
           \psi(\lambda) \cdot \psi^*(\lambda) :
    \frac{d\lambda}{2 \pi i},                        \tag 34
$$
where the integral is along a circle $|\lambda| = \text{const.}$,
corresponds to the action of $\hbar^{-1}W^{(\ell)}_n(\hbar)$
on $\tau(\hbar,t)$, from which we can derive $w^{(\ell)}_n F$
as described above. This can be generalized to fermion bilinear
forms of the form
$$
    \calO_A(\hbar) =
    \oint : A\left(\lambda,\hbar\frac{\rd}{\rd \lambda}\right)
                 \psi(\lambda) \cdot \psi^*(\lambda):
    \frac{d\lambda}{2 \pi i}.                        \tag 35
$$
They generate $W_{1+\infty}$ symmetries $\hbar^{-1}W_A(\hbar)$ that
contract to $w_{1+\infty}$ symmetries $w_A F$ ($ =-\delta_A F$
in the notation of our previous work [10]) of the dispersionless
KP hierarchy. If one can exponentiate (i.e., integrate) these
infinitesimal symmetries, finite symmetries thus obtained on the
space of solutions should generate nontrivial solutions out of the
trivial solution $\tau(\hbar,t)=1$. This is indeed achieved by
inserting a Clifford operator exponentiating a fermion bilinear form
$\calO_A(\hbar)$; $g(\hbar)$ should be such a Clifford operator.
Actually, to ensure correct asymptotic behavior of $\tau(\hbar,t)$
in the limit of $\hbar \to 0$,  the exponent should be
$\hbar^{-1}\calO_A(\hbar)$ rather than $\calO_A(\hbar)$, just
like $\hbar^{-1}H(t)$ in the boost operator of time evolutions.
We are thus led to the following ansatz:
$$
    g(\hbar) = \exp \hbar^{-1} \calO_A(\hbar).      \tag 36
$$

The validity of this ansatz has been checked by several different
methods. Further, these solutions, depending on an arbitrary
function $A(\lambda,\mu)$, actually give general solutions of the
dispersionless KP hierarchy. This is a consequence of our previous
construction of $w_{1+\infty}$ symmetries based on a kind of
Riemann-Hilbert problem [10].

We have thus seen several aspects of dispersionless hierarchies
from the standpoint of quasi-classical limit, taking the
dispersionless KP hierarchy as an example. These results can be
extended to other hierarchies, such as the modified KP and Toda
hierarchies. A new characteristic of these two hierarchies is
the presence of a discrete variable, say $\nu \in \bfZ$, other
than continuous ones like $t_n$'s. In quasi-classical limit,
such a variable has to be rescaled as $\hbar \nu = s$,
$s$ being a continuous variable.

Our results in this and previous papers suggest a deep link of
dispersionless hierarchies with recent various approaches to
$c=1$ matrix models [24] and their ``quantum fluid'' picture in
quasi-classical limit [25]. In particular, our field theoretical
tools are technically very close to the free fermion formalism of
Das et al., though dealing with apparently different models.
Elucidating these possible relations is an intriguing issue,
which might lead to a similar approach to 4D self-dual gravity.

%%%%% references %%%%%%%%%%%%%%%%%%%%%%%%%%%%%%%%%%%%%%%%%%%%%%%%%
\title References
\endtitle

\item{[1]}
Lebedev, D., and Manin, Yu.,
%%% Conservation Laws and Lax Representation
%%% on Benny's Long Wave Equations,
Phys.Lett. 74A (1979), 154--156.
\item{}
Kodama, Y.,
%%% A method for solving the dispersionless KP equation and
%%% its exact solutions,
Phys. Lett. 129A (1988), 223-226;
%%% Solutions of the dispersionless Toda equation,
Phys. Lett. 147A (1990), 477-482.
\item{}
Kodama, Y., and Gibbons, J.,
%%% A method for solving the dispersionless KP hierarchy and
%%% its exact solutions, II,
Phys. Lett. 135A (1989), 167-170;

\item{[2]}
Krichever, I.M.,
%%% The dispersionless Lax equations and topological minimal models,
Commun. Math. Phys. 143 (1991), 415-426;
%%% The $\tau$-function of the universal Whitham hierarchy,
%%% matrix models and topological field theories,
LPTENS-92/18 (May, 1992).

\item{[3]}
Dubrovin, B.A.,
%%% Hamiltonian formalism of Whitham-type hierarchies
%%% and topological Landau-Ginsburg models,
Commun. Math. Phys. 145 (1992), 195-207.

\item{[4]}
Dijkgraaf, R., Verlinde, H., and Verlinde, E.,
%%% Topological strings in $d<1$,
Nucl. Phys. B352 (1991), 59-86.

\item{[5]}
Blok, B., and Varchenko, A.,
%%% Topological conformal field theories and flat coordinates,
Int. J. Mod. Phys. A7 (1992), 1467-1490.

\item{[6]}
Bakas, I.,
%%% The structure of the $W_\infty$ algebra,
Commun. Math. Phys. 134 (1990), 487-508.

\item{[7]}
Park, Q-Han,
%%% Extended conformal symmetries in real heavens,
Phys. Lett. 236B (1990), 429-432.

\item{[8]}
Saveliev, M.V., and Vershik, A.M.,
%%% Continual analogues of contragredient Lie algebras,
Commun. Math. Phys. 126 (1989), 367-378;

\item{[9]}
Park, Q-Han,
%%% Self-dual gravity as a large-$N$ limit
%%% of the 2D non-linear sigma model,
Phys. Lett. 238B (1990), 287-290;
%%% 2-D sigma model approach to 4-D instantons,
Int. J. Mod. Phys. A7 (1992), 1415-1448.

\item{[10]}
Takasaki, K., and Takebe, T.,
%%% SDiff(2) Toda equation -- hierarchy, tau function and symmetries,
Lett. Math. Phys. 23 (1991), 205-214;
%%% SDiff(2) KP hierarchy,
in: Proceesings of RIMS Research Project 1991 ``Infinite Analysis,"
Int. J. Mod. Phys. A7, Suppl. 1B (1992), 889-922.

\item{[11]}
For an overview of W-infinity algebras, see:
Sezgin, E.,
%%% Aspects of $W_\infty$ symmetry,
Texas A\& M preprint CTP-TAMU-9/91, IC/91/206;
%%% Area-preserving diffeomorphisms, $w_\infty$ algebras
%%% and $w_\infty$ gravity,
Texas A\& M preprint CTP-TAMU-13/92.

\item{[12]}
Boyer, C.P., and Plebanski, J.F.,
%%% An infinite hierarchy of conservation laws
%%% and nonlinear superposition principles for
%%% self-dual Einstein spaces,
J. Math. Phys. 26 (1985), 229-234.

\item{[13]}
Takasaki, K.,
%%%% Symmetries of hyper-K\"ahler
%%% (or Poisson gauge field) hierarchy,
J. Math. Phys. 31 (1990), 1877-1888.

\item{[14]}
Dijkgraaf, R., Verlinde, E., and Verlinde, H.,
%%% Loop equations and Virasoro constraints in
%%% non-perturbative 2d quantum gravity,
Nucl. Phys. B348 (1991), 435-456.

\item{[15]}
Fukuma, M., Kawai, H., and Nakayama, R.,
%%% Continuum Schwinger-Dyson equations and universal
%%% structures in two-dimensional quantum gravity,
Int. J. Mod. Phys. A6 (1991), 1385-1406;
%% Infinite dimensional Grassmannian structure of
%% two dimensional string theory,
Commun. Math. Phys. 143 (1991), 371-403.

\item{[16]}
Awada, M., and Sin, S.J.,
%%% The string difference equation of the $D=1$ matrix model
%%% and $W_{1+\infty}$ symmetry of the KP hierarchy,
Int. J. Mod. Phys. A7 (1992), 4791-4802.

\item{[17]}
Yoneya, T.,
%%% Toward a canonical formalism of non-perturbative
%%% two-dimensional gravity,
Commun. Math. Phys. 144 (1992), 623-639.

\item{[18]}
Adler, M., and van Moerbeke, P.,
%%% A matrix integral solution to two-dimensional $W_p$-gravity,
Commun. Math. Phys. 147 (1992), 25-56.

\item{[19]}
Sato, M., and Sato, Y.,
%%% Soliton equations as dynamical systems
%%% in an infinite dimensional Grassmann manifold,
in: {\it Nonlinear Partial Differential Equations
in Applied Sciences}
(North-Holland, Amsterdam, and Kinokuniya, Tokyo, 1982).

\item{[20]}
Segal, G., and Wilson, G.,
%%% Loop groups and equations of KdV type,
Publ. IHES 61 (1985), 5-65.

\item{[21]}
Date, E., Kashiwara, M., Jimbo, M., and Miwa, T.,
%%% Transformation groups for soliton equations,
in: {\it Nonlinear Integrable Systems ---
Classical Theory and Quantum Theory}
(World Scientific, Singapore, 1983).

\item{[22]}
Orlov, A.Yu.,
%%% Vertex operators, $\bar{\partial}$-problems, symmetries, variational
%% indentities and Hamiltonian formalism for $2+1$ integrable systems,
in: {\it Plasma Theory and Nonlinear and
Turbulent Processes in Physics\/}
(World Scientific, Singapore, 1988).
\item{}
Grinevich, P.G., and Orlov, A.Yu.,
%%% Virasoro action on Riemann surfaces, Grassmannians,
%%% $\det\bar{\partial}_j$ and Segal Wilson $\tau$ function,
in: {\it Problems of Modern Quantum Field Theory\/}
(Springer-Verlag, 1989).

\item{[23]}
Itzykson, C., Parisi, G. and Zuber, J.B.,
%%% Planar Diagrams.
Commun. Math. Phys. 59 (1978), 35-51.

\item{[24]}
Das, S.R., and Jevicki, A.,
%%% String Field Theory and Physical Interpretation of d=1 Strings
Mod. Phys. Lett. A5 (1990), 1639-1650.

\item{}
Avan, J., and Jevicki, A.,
%%% Classical Integrability and Higher Symmetries
%%% of Collective String Field Theory
Phys. Lett. B266 (1991), 35-41.

\item{}
Witten, E.,
%%% Ground ring of two dimensional string theory
Nucl. Phys. B373 (1992), 187-213.

\item{}
Klebanov, I.R., and Polyakov, A.M.,
%%% Interaction of discrete states in two-dimensional string theory
Mod. Phys. Lett. A6 (1991), 3273-3281.

\item{}
Das, S.R., Dhar, A., Mandal, G., and Wadia, S.R.,
%%% Gauge theory formulation of the $c=1$ matrix model:
%%% symmetries and discrete states,
Int. J. Mod. Phys. A7 (1992), 5165-5195.

\item{[25]}
Polchinski, J.,
%%% Classical Limit of 1+1 Dimensional String Theory
Nucl. Phys. B362 (1991), 125-140.

\item{}
Dhar, A., Mandal, G., and Wadia, S.R.,
%%% Classical Fermi fluid and geometrical action for $c=1$,
IASSNS-HEP-91/89 (March, 1992).

\item{}
Iso, S., Kalabali, D., Sakita, B.,
%%% One dimensional fermions as two-dimensional droplets via
%%% Chern-Simons theory
CCNY-HEP-92/1 (January, 1992).

\bye